# Image charge interaction correction in charged-defect calculation

Zhao-Jun Suo[1,2], Jun-Wei Luo[1,2,3]*, Shu-Shen Li[1,2,3], and Lin-Wang Wang[4]*

[1]*State Key Laboratory of Superlattices and Microstructures, Institute of Semiconductors, Chinese Academy of Sciences, Beijing 100083, China*
[2]*Center of Materials Science and Optoelectronics Engineering, University of Chinese Academy of Sciences, Beijing 100049, China*
[3]*Beijing Academy of Quantum Information Sciences, Beijing 100193, China*
[4]*Materials Science Division, Lawrence Berkeley National Laboratory, Berkeley, California 94720, United States*

*Email: jwluo@semi.ac.cn; lwwang@lbl.gov

## Abstract

Charged-defect calculation using a periodic supercell is a significant class of problems in solid state physics. However, the finite supercell size induces an undesirable long-range image charge Coulomb interaction. Although a variety of methods have been proposed to eliminate such image Coulomb interaction, most of the previous schemes are based on a rough approximation of the defect charge screening. In this work, we present a rigorous derivation of the image charge interaction with a new defect screening model where the use of bulk macroscopic dielectric constant can be avoided. We have verified this approach in comparison with a widely used approach for 12 different defects. Our correction scheme offers a much faster convergence concerning the supercell size for cases with considerable image charge interactions. In those cases, we also found that the nonlinear dielectric screening might play an important role. The proposed new defect screening model will also shed new light on understanding the defect screening properties and can be applied to other defect systems.

## I. INTRODUCTION

Defects in semiconductors play an essential role in controlling the semiconductor properties [1-4]. It has been a long history of calculating defect properties, including structures, formation energies, and transition levels, using density functional theory (DFT) methods [5,6]. Although well-established defect computational procedures have been developed, first-principles calculations of such defect properties remain challenges. Besides the fundamental problems related to DFT functional itself (e.g., the well-known underestimation of the band gap), there are also practical issues related to the utilization of a finite-size supercell [7-9]. Specifically, one can ideally calculate defect properties in an infinite system, but, in practice, a finite-size supercell is always used for saving computational cost. In scenarios such as shallow donors or acceptors, the defect wave function is very extended and spreads over a large space so that a large supercell must be utilized [10,11]. Whereas, for a relatively deep defect, the defect wave function is rather localized and thus can be contained in a small supercell. As long as the supercell is larger than the spreading of the defect wave function and is converged in elastic relaxation, the result of finite supercell approaches to that of the infinite system very quickly for charge-neutral defects where the total charge around the defect site is zero. However, if the defect is charged, the convergence is slow due to the long-range image-image charge interaction resulting from the utilization of a finite supercell in the periodic boundary condition. Here, the electrostatic potential is calculated, relying on a uniform (jellium) background charge that compensates the defect charge in the system [12]. Due to the slow convergence of image charge interaction, it is impractical to use an increasing larger supercell trying to reach the converged value [13]. Then it is necessary to provide an image charge correction to make the convergence faster to the infinite supercell result or even new scheme beyond the jellium model. Although different formulas and methods [14-23] have been proposed, uncertainties exist in the applications of these formulas. These uncertainties result in scattered data presented in the literature for the calculated formation energies and transition levels, leading to controversies on defect calculations. For example, the formation energies of O vacancy in ZnO using different schemes and procedures to account for LDA/GGA deficiencies and finite-size effects differed a lot [7]. Specifically, Yu-Ning Wu et al. claimed the transition level for $V_O^{2+}$ is $E_V$ +1.5 eV [24]. recently, however, Wei Chen, Alfredo Pasquarello [25] and Hui-Xiong Deng, Su-Huai Wei [26] gave about 1 eV higher results and commented that Yu-Ning Wu's work might deal with image charge interaction improperly. It is thus highly desired to have a more rigorous and robust method to correct the image charge interaction.

One of the early works on image charge interaction correction is done by Makov and Payne (MP method) [15]. They used a Madelung interaction for a point charge in a lattice and a second moment of the defect charge to capture the finite defect charge size to describe the image interactions as $1/L$ and $1/L^3$ terms ($L$ is the size of the supercell), respectively. Their formula has been carefully tested and revised by Lany and Zunger (LZ) [7,8] to provide a better convergence. Freysoldt, Neugebauer, and Van de Walle (FNV) [16,17] alternatively proposed a new approach by assuming the shape of the defect charge as a spherical model with model parameters obtained by fitting the defect wave functions and comparing the calculated potential in the supercell. The obtained explicit charge model is then used to correct the image charge interaction. The FNV is currently widely used as a standard approach for calculations of charged defects. In order to treat defects in non-cubic systems, Samuel T. Murphy and Nicholas2017 D. M. Hine modified the charged-defect screening model by considering the anisotropic dielectric properties [19]. Similarly, Yu Kumagail and Fumiyasu Oba extended the FNV approach to consider the dielectric tensor for anisotropic materials [20]. All these works have contributed significantly to this field, but the variations of approaches complicate the calculations and imply uncertainties. Furthermore, all these methods use macroscopic dielectric constants or properties to represent the screening of the defect charge. However, the screening of the defect charge could be quite different from the bulk macroscopic case due to the small spread of the defect charge, local field effects, as well as nonlinear screenings. Besides, because of the finite size of the supercell, a compensating jellium background charge has been added for the polarization charge to make the total response charge equal zero. All of these aspects make the situation more complicated.

In this work, we derive a correction formula rigorously for image charge interaction based on a defect screening model. Unlike previous approaches, where analytical formulas with adjustable parameters are used to describe the image charge interaction or defect charge profile, we will base our correction directly on numerical results contained in calculations. In other words, we carry out relatively cheap, but numerical post-processing calculations to provide the image charge corrections. We believe such a numerical approach provides higher robustness, especially for cases with an elongated (instead of spherical) defect charge, or for cases when nonlinear screening is important. We use the self-consistent field (SCF) calculated charge density instead of the macroscopic dielectric constant, considering it already contains the information for the screening. This provides an alternative way to consider the image charge correction problem compared to the previous analytical formulas. It is demonstrated that our method can provide much faster convergence than previous methods for cases where the image charge interaction is considerable.

It is worth noting that the screening of the defect charge comes from both electronic and ionic parts. This is also a source of controversy in using analytical formulas presented in previous correction methods. For example, Peter Deák et al. compared results of using $\varepsilon_\infty = 3.55$ (the high-frequency dielectric constant) or $\varepsilon_0 = 10$ (the static dielectric constant) in calculating the adiabatic charge transition levels in β-$Ga_2O_3$, and got that using $\varepsilon_\infty$ could yield better agreement with experimental data whereas using $\varepsilon_0$ would underestimate by 1 eV [27]. They inferred that the actual screening in the supercell should be described by a dielectric constant between $\varepsilon_0$ and $\varepsilon_\infty$ [27]. However, conceptually, either $\varepsilon_0$ should be used when the atomic positions are relaxed or $\varepsilon_\infty$ should be applied when the atomic positions are held fixed (vertical ionization) after ionization [8,27,28]. In this paper, we primarily deal with the situation where only the electronic screening effect (i.e., $\varepsilon_\infty$) is applied. Although the formula we derived can also be applied to ionic screening, we leave it to future studies to establish our defect screening model for the ionic screening effect. The challenge is that one needs to fully relax the atomic positions for different size supercell calculations, rendering it more challenging than an SCF electronic structure calculation.

This paper is organized as follows. In Sec. II, we describe our correction scheme, including the generally applied formation energy formulas (Sec. II A), C-NS correction in the non-SCF unscreened situation and C-AP using $\varepsilon_\infty$ (Sec. II B), defect state screening model obtained through SCF calculation of neutral and charged systems (Sec. II C), a precise theory of image charge correction C-SC for SCF screened situation (Sec. II D), comparison with FNV (Sec. II E) and LZ (Sec. II F), and a definition of the effective dielectric constant (Sec. II G). Then we apply our method C-AP and C-SC to calculate the energies for different defects and discuss the results in Sec. III Finally, we conclude our work in Sec. IV.

## II. CHARGED-DEFECT ENERGY CORRECTION

### A. Formation energy and the correction

In a supercell method, the formation energy of a charged defect is often described as [6,20,28-30]

$$\Delta H_f(\alpha, q) = \{E(\alpha, q) + E_{corr}(\alpha, q)\} - E(host) - \sum_i n_i \mu_i + q(\varepsilon_F + \varepsilon_{VBM} + \Delta v). \quad (1)$$

Here $E(\alpha, q)$ is the total energy of a supercell containing a defect $\alpha$ with charge $-q$ ($q$ electron), and $E(host)$ is the total energy of pristine bulk crystal with the same supercell except for the defect. In order to form the defect, $n_i$ atoms with chemical potential $\mu_i$ are added ($n_i > 0$) to or removed ($n_i < 0$) from the supercell. The Fermi energy $\varepsilon_F$ is referenced to the energy of valence band maximum (VBM) $\varepsilon_{\text{VBM}}$ of the host (e.g., $\varepsilon_F = 0$ when the Fermi energy is at the VBM). Both $E_{corr}(\alpha, q)$ and $q\Delta v$ are charge correction terms due to the usage of the finite-size supercell. They scale as the inverse of supercell size and the inverse of the supercell volume, respectively. Thus, they tend to zero as the supercell size towards infinity. Roughly speaking, $E_{corr}(\alpha, q)$ corresponds to an image charge interaction term, while $q\Delta v$ represents a potential alignment $\Delta v$ (PA) term between the pristine bulk crystal and the charge-neutral system at a place far away from the defect. While the first term is proportional to $q^2$, the second term is proportional to $q$. Therefore, for a neutral defect, these terms are zero. Together, we can call them the charged-defect energy correction,

$$E_C(\alpha, q) = E_{corr}(\alpha, q) + q\Delta v. \tag{2}$$

This $E_C(\alpha, q)$ is used to get the infinite system results from finite supercell calculations. The PA correction term is well understood, and there are several approaches to deal with the image charge interaction term, $E_{corr}(\alpha, q)$. Although derivations of $E_{corr}(\alpha, q)$ are frequently based on non-self-consistent field (non-SCF) approximations, there is no direct derivation of the $E_{corr}(\alpha, q)$ term for SCF calculations and just an intuitively approximation through employing a dielectric screening constant. Due to the lack of derivation, many treatments are only guided by intuition, which can lead to controversy.

To provide a rigorous derivation and an expression for $E_{corr}(\alpha, q)$, we propose a defect charge screening model, which offers a way to treat the screening of the defect charge from SCF calculations. This defect screening model is tested and validated using our numerical calculations. Finally, our $E_{corr}(\alpha, q)$ expression is applied to actual defect calculations. We find that, compared to previous models, our new formalism can provide faster convergence, especially for cases associated with multiple charge states ($q > 1$), the value of the correction term is large, and nonlinear screening effect might exist.

In this paper, however, we only deal with the screening effect due to wave function SCF treatment (e.g., the electronic structure part), and ignore the ionic screening component. In other words, from small supercell to large supercell, we will fix the atomic positions as in the original small supercell, while using the pristine crystal to fill the rest of the large supercell. The atomic relaxation near the defect induced by the change of defect charge has been included, although it is just for the small supercell system. For polar crystal, the ionic displacement for atoms far away from the defect can still provide further screening effect, which reduces the image interaction. Our final formula is likely also applicable to ionic screening as long as the nuclear charges are treated as part of the total charge (e.g., using a Gaussian broadening to represent the nuclear charge). Nevertheless, the atomic relaxation for a few hundred atoms is not an easy task, and the Gaussian broadening presents some additional technical issues. The defect charge screening model has to be tested separately for the ionic screening. Thus, in the current calculations, we will neglect the ionic screening effects and leave such effects in future investigations.

**B. Non-SCF unscreened situation**

We start with the most straightforward situation in which the change of electronic wave functions caused by the occupation of defect state orbitals is ignored. Therefore, only the electrostatic interaction is considered. We assume the defect is in charge state $-q$ ($q$ is the number of electrons) with a wave function $\varphi'_d$. To simplify the notation in the following derivation, we will define $\varphi_d = \sqrt{|q|}\varphi'_d$, and assume the defect charge density is $\rho_d(r)$ (with its spatial integral being $q$). Then we will have

$$\rho_d = \varphi_d^2, \tag{3}$$

$$E_0 = \frac{1}{2}\int (\rho_0(r) + \rho_{ion}(r))\frac{1}{|r-r'|}(\rho_0(r') + \rho_{ion}(r'))d^3r d^3r', \tag{4}$$

$$E_q = \frac{1}{2}\int (\rho_d(r) + \rho_0(r) + \rho_{ion}(r))\frac{1}{|r-r'|}(\rho_d(r') + \rho_0(r') + \rho_{ion}(r'))d^3r d^3r', \tag{5}$$

where $E_0$ and $E_q$ are the electrostatic energy of the supercell with the neutral and charged defect, respectively. $\rho_0$ is the total electron density of the neutral charge system, and $\rho_{ion}$ is the nuclear charge. So, the electrostatic energy deviation induced by the charged defect is

$$\Delta E^N = E_q - E_0$$
$$= \frac{1}{2}\int \rho_d(r)\frac{1}{|r-r'|}\rho_d(r')d^3rd^3r' + \int \rho_d(r)\frac{1}{|r-r'|}(\rho_0(r')+\rho_{ion}(r'))d^3rd^3r' \quad (6)$$
$$= \frac{1}{2}\int \rho_d(r)\frac{1}{|r-r'|}\rho_d(r')d^3rd^3r' + \int \rho_d(r)V_{tot}(r)d^3r.$$

This $\Delta E^N$ will depend on the supercell size N due to the Hartree interaction integral in a finite supercell. What we intend to do is to provide a correction term so that we can obtain $\Delta E^\infty$ for the infinite system from the finite supercell calculation of $\Delta E^N$. Because $E_0$ does not need any correction, and all the correction is for the q-charged defect formation energy of Eqs. (1) and (2). Note, in the Hartree calculation for the finite supercell, we have assumed a constant compensating background charge. In other words, we have required that $V_{tot}(G=0) = 0$ with reciprocal vector G. As a result, even for neutral defect system, we have $V_{tot}(r) = V_{tot}^\infty(r) + \Delta V_N$ ($\Delta V_N$ will be the PA term at supercell size N). We then have

$$E_{C-NS}^N = \Delta E^\infty - \Delta E^N$$
$$= \frac{1}{2}\int_\infty \rho_d(r)\frac{1}{|r-r'|}\rho_d(r')d^3rd^3r' - \frac{1}{2}\int_{P,N}\rho_d(r)\frac{1}{|r-r'|}\rho_d(r')d^3rd^3r' \quad (7)$$
$$-q\Delta V_N,$$

where $\int_{P,N}$ means the Coulomb integration within an N sized periodic supercell with a constant compensating background charge (i.e., it can be done in reciprocal space by setting the $G=0$ component to zero), while $\int_\infty$ denotes the integral in an infinite supercell under an open boundary condition for the Poisson equation (see the supplemental information, section 9 for details). C-NS (correction for no screening effect) method is implemented for the non-self-consistent field (non-SCF) calculated result of the charged-defect system, where all the wave functions are from neutral state calculation but with different electron occupations. In the non-SCF calculation for charged-defect system, the defect wave functions together with its charge density are fixed, and there are no SCF iterations. Therefore, the screening response to defect ionization is not included in the non-SCF calculated result.

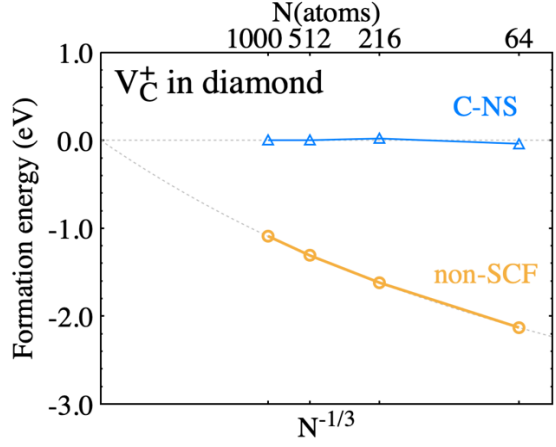

FIG.1. C-NS correction on non-SCF calculated energies of $V_C^+$ in diamond supercells. All the formation energies are referenced to C-NS corrected formation energy (which is set to zero) at the largest supercell (containing 1000 atoms). The numbers at the top of the panel denote the numbers of atoms in different supercells. And the gray dashed lines are just for the guide of the eye.

Fig. 1 shows the reliability of C-NS in Eq. (7) under non-SCF calculation results. For $V_C^+$ in different diamond supercells, the non-SCF calculated energies vary significantly with large amplitudes. After C-NS correction, the formation energies get converged quickly. The vast energy differences between non-SCF and C-NS indicate the strong image charge Coulomb interaction in this unscreened case. On the other hand, the interaction energy will decrease if the screening effect is taken into consideration. Very often, to represent the screening effect, a macroscopic dielectric constant $\varepsilon$ is added in front of the Coulomb interaction term. Then we get

$$E_{C-AP}^N = \frac{1}{2\varepsilon}\int_\infty \rho_d(r)\frac{1}{|r-r'|}\rho_d(r')d^3rd^3r' - \frac{1}{2\varepsilon}\int_{P,N} \rho_d(r)\frac{1}{|r-r'|}\rho_d(r')d^3rd^3r' \\ -q\Delta V. \quad (8)$$

This is nevertheless only an approximation since the macroscopic dielectric constant can only describe the screening effect of slowly varying perturbation potential, not the defect charge, where rapidly varying local field effect is crucial. Furthermore, as we will show later, there are examples where the nonlinear screening effect is considerable, which can make the effective dielectric constant charge dependent.

**C. Screening model for the charged defect**

The non-SCF unscreened situation is simple, and its finite-size correction is straightforward. Whereas, the SCF situation with screening is much more complicated. In order to yield a useful formula, we first propose and test a defect charge screening model. This screening model will be used in subsequent derivations. We note that this model can include not only linear screening but also nonlinear screening effects as we do not assume the proportionality of the defect charge $q$. We first define $\rho_{d,sc}^{N}(r)$ as the SCF charge density difference between the supercell with a charged defect $\rho_{q,sc}^{N}(r)$ and the supercell with a neutral defect $\rho_{0,sc}^{N}(r)$,

$$\rho_{d,sc}^{N}(r) = \rho_{q,sc}^{N}(r) - \rho_{0,sc}^{N}(r). \tag{9}$$

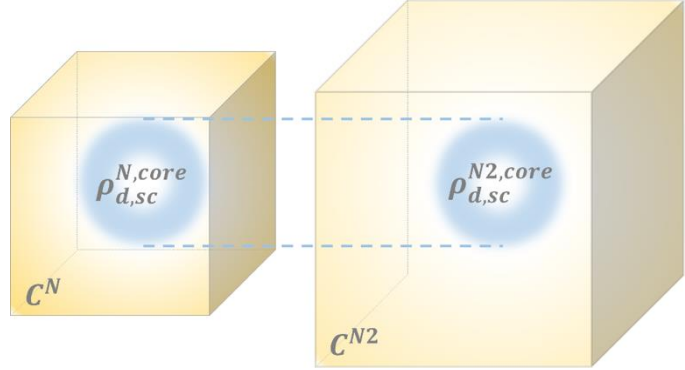

FIG.2. Supercell scaling of charge densities. In each supercell, the screened defect charge density is separated into $\rho_{d,sc}^{N,core}(r)$ and $C^N(r)$. The core part $\rho_{d,sc}^{N,core}(r)$ remains the same in both N and N2 supercells: $\rho_{d,sc}^{N,core}(r) = \rho_{d,sc}^{N2,core}(r)$. The background part $C^N(r)$ in yellow shallows in supercell N2 due to their inverse dependence on supercell volume.

One can regard $\rho_{d,sc}^{N}(r)$ as the screened defect charge of the unscreened bare defect charge density $\rho_d(r)$ plus all the possible background compensating charges. Note, in the above formalism, to concern only the electronic screening effect and neglect the ionic screening, we have used the same atomic positions for both charged and neutral defects, although atomic positions may response to defect charging and be the charged-defect positions (even for the neutral defect calculation) for the atomic coordinates near the defect.

Here we propose a model to describe the behavior of $\rho_{d,sc}^{N}(r)$, which helps to explain how the bare defect charge is screened. We first separate $\rho_{d,sc}^{N}(r)$ into a core part $\rho_{d,sc}^{N,core}(r)$ and a background part $C^N(r)$ as schematically shown in Fig. 2. We suppose that the core part remains almost the same in different supercells as long as the supercell size N is large enough to contain $\rho_{d,sc}^{N,core}(r)$. One can consider $\rho_{d,sc}^{N,core}(r)$ as the screened $\rho_d(r)$ in the infinite system. $C^N(r)$ is extended throughout the finite supercell and is responsible for the zero total polarization charge. The majority of the background charge $C^N(r)$ away from the defect is a constant. Since the total charge of this compensating charge is the same when varying the supercell size, thus for two supercells with volumes $\Omega^N$ and $\Omega^{N2}$, we have (for $r$ within the domain of the smaller supercell)

$$\Omega^N \cdot C^N(r) = \Omega^{N2} \cdot C^{N2}(r). \tag{10}$$

Since $\rho_{d,sc}^{N,core}(r)$ is independent of supercell size N and can be written as $\rho_{d,sc}^{\infty,core}(r)$. Hence,

$$\rho_{d,sc}^N(r) = \rho_{d,sc}^{\infty,core}(r) + C^N(r). \tag{11}$$

Through Eqs. (9)-(11), $C^N(r)$ can be acquired from $\rho_{d,sc}^N(r)$ and $\rho_{d,sc}^{N2}(r)$ of two supercells N and N2:

$$C^N(r) = \frac{\Omega^{N2}}{\Omega^{N2} - \Omega^N} \left( \rho_{d,sc}^N(r) - \rho_{d,sc}^{N2}(r) \right) \bigg|_{r \in \Omega^N}. \tag{12}$$

The $C^N(r)$ from Eq. (12) can also be used to obtain $\rho_{d,sc}^{N,core}(r)$ through

$$\rho_{d,sc}^{N,core}(r) = \rho_{d,sc}^N(r) - C^N(r). \tag{13}$$

This $\rho_{d,sc}^{N,core}(r)$ can be used to check whether it is indeed independent of supercell size N, as well as the N-independence of $C^N \cdot \Omega^N$. The results are shown in Fig. 3 with their radial integrations. As we can see, they are both N-independent. The flat tail in the integration of $C^N \cdot \Omega^N$ after $r$ reaching the supercell edge is the feature of the cubic supercell. However, the collapsing of different curves into a single one indicates they are the same before the supercell edge is reached. In Fig. 3(B) and 3(D), we also compared the integration curve of $C^N \cdot \Omega^N$ with that of a homogeneous charge density (of equal total charges) for all the supercells. They are both very close, indicating the background change $C^N$ can be approximated as a homogeneous charge (variation nearby the defect is unavoidable but does not account for the majority of its charge) as we assumed in the derivation of Eq. (10). Subsequently, Fig. 3 validates our defect charge screening model. We find similar situations for all other investigated defects. This model will play an essential role in evaluating the electrostatic charge of the defect state. Instead of using a macroscopic dielectric constant, here we will use the SCF calculated charge density $\rho_{d,sc}^N(r)$ to figure out the dielectric screening. This is natural since the dielectric screening effects (including small size effect, local field effect, and nonlinear screening effect) have already been captured in the SCF calculation. There is no necessity for additional calculations.

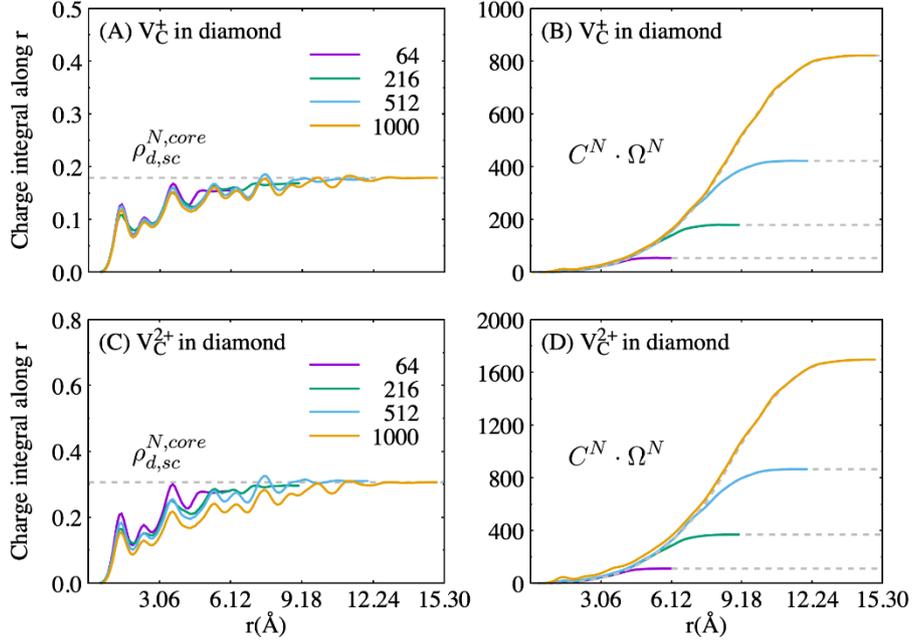

FIG.3. Integral of defect charge density in supercells containing 64 to 1000 atoms. (A) and (C) are core part $\rho_{d,sc}^{N,core}(r)$ integrating along the radial direction of $V_C^+$ and $V_C^{2+}$ in different supercells. (B) and (D) are $C^N \cdot \Omega^N$ integration of $V_C^+$ and $V_C^{2+}$. The supercell volume $\Omega^N$ is replaced with the number of atoms in the supercell for simplicity. The dashed lines in (B) and (D) are the integrated results of uniform charge densityies with the total charges equaling those of $C^N \cdot \Omega^N$. They show that the $C^N$ can be well approximated as a uniform charge density.

### D. SCF screened defect and energy correction

Based on the above screening model for the screened defect charge density, we can now give an expression for the electrostatic interaction and its related energies and provide a correction formula from calculations of a finite supercell to obtain the infinity system results. However, unlike the case of non-SCF calculations, where only the electrostatic interaction energy related to the defect charge is presented in Eqs. (3)-(7) (since the other energies, including the kinetic energy, nonlocal potential energy, and the electrostatic interactions for the rest of the charge, are fixed during a non-SCF calculation), one has to include all the energy terms in the SCF calculation. Let $E_{0,scf}^N$ be the total energy for the N-sized periodic supercell with a neutral defect, and its total electron charge density is $\rho_{0,sc}^N(r)$. $E_{0,scf}^N$ includes kinetic energy $T_0$, electron-electron Coulomb energy, electron-ion Coulomb energy, the nonlocal pseudopotential energy $E_{nl}$, and the exchange-correlation energy $E_{xc}$:

$$E_{0,scf}^N = T_0 + \frac{1}{2}\int \rho_{0,sc}^N(r)\frac{1}{|r-r'|}\rho_{0,sc}^N(r')d^3r d^3r' + \int \rho_{0,sc}^N V_{ion}(r)d^3r + E_{nl} \\ + \int E_{xc}(\rho_{0,sc}^N)d^3r. \tag{14}$$

When the defect is ionized with charge $q$, the initial (non-SCF) charge density is $\rho_{0,sc}^N + \rho_d$, where $\rho_d$ is given in Eq. (3). After SCF iterations, the total electron charge density finally converges to $\rho_{q,sc}^N = \rho_{0,sc}^N + \rho_{d,sc}^N = \rho^N + \rho_d$ (the first part is just Eq. (9), the second part is a definition of $\rho^N$). However, from now on, $\rho_d = \varphi_d^2$, but $\varphi_d$ should be understood as the defect state wave function from SCF calculations with q-charge, instead of neutral charge calculation (although, our later test shows, the effect of this relaxation on $\varphi_d$ is rather negligible as far as the image correction term is concerned). We can further define the polarization charge density (the charge responsible for the screening) as

$$\Delta\rho^N = \rho^N - \rho_{0,sc}^N = \rho_{d,sc}^N - \rho_d. \tag{15}$$

Once again, $\rho^N = \rho_{q,sc}^N - \rho_d$ is the electron density of the charged-defect system subtracted the defect charge density $\rho_d$. Now, we can write down the expression for the SCF total energy $E_q^N$. The SCF wave functions are $\{\varphi_{i,scf}^N, \varphi_d\}$, where $\varphi_d$ is the defect state wave function from SCF calculation as discussed above and is independent of the supercell size N, and then $\rho_d = \varphi_d^2$ and $\rho^N = \sum_i |\varphi_{i,scf}^N|^2$. Plugging the SCF solutions $\{\varphi_{i,scf}^N, \varphi_d\}$ into a DFT total energy formula, after some simple derivations, we have the SCF total energy of the q-charged defect system as (keep the energy in the second-order expansion of $\rho_d$)

$$E_q^N = E_{tot}^N[\rho^N\{\varphi_{i,scf}^N\}] + \int \rho_d V_{tot}(\rho^N\{\varphi_{i,scf}^N\}, r)d^3r \\ + (T + E_{nl})(\varphi_d) + \frac{1}{2}\int \rho_d(r)\frac{1}{|r-r'|}\rho_d(r')d^3r d^3r' + \frac{1}{2}\int \frac{\delta}{\delta\rho}V_{xc}[\rho]\bigg|_{\rho=\rho^N}\rho_d^2(r)d^3r, \tag{16}$$

$$E_{tot}^N[\rho^N\{\varphi_{i,scf}^N\}] = (T+E_{nl})[\varphi_{i,scf}^N] + \frac{1}{2}\int \rho^N(r)\frac{1}{|r-r'|}\rho^N(r')d^3rd^3r' \tag{17}$$
$$+ \int \rho^N V_{ion}(r)d^3r + \int E_{xc}(\rho^N)d^3r.$$

Note that $E_{tot}^N[\rho^N\{\varphi_{i,scf}^N\}]$ is just the total energy expression for the neutral system. If $\{\varphi_{i,scf}^N\}$ are allowed to change variationally, the minimum solution of $E_{tot}^N[\rho^N\{\varphi_i^N\}]$ is $E_{0,scf}^N$ corresponding to the SCF total energy of the neutral system. On the other hand, the minimum solution of $E_q^N$ in Eq. (16) related to the variational change of $\{\varphi_{i,scf}^N\}$ (e.g., the SCF solution) can be considered as a response to $E_{tot}^N[\rho^N\{\varphi_{i,scf}^N\}]$ of the system under an external perturbation caused by the second term in Eq. (16). In this regard, we can rewrite the second line of Eq. (16) as $E_{fix}[\rho_d]$, which is a fixed energy term during $\{\varphi_{i,scf}^N\}$ variation and does not play any role. Following a DFT perturbation theory treatment [31], the second term in Eq. (16) can be further approximated (to the second order) as

$$\int \frac{\delta}{\delta\rho(r')}V_{tot}[\rho](r)\bigg|_{\rho=\rho_{0,sc}^N} \Delta\rho^N(r')\rho_d(r)d^3r'd^3r + \int \rho_d V_{tot}(\rho_{0,sc}^N)d^3r. \tag{18}$$

Here, $\Delta\rho^N = \rho^N - \rho_{0,sc}^N$ is defined in Eq. (15). Furthermore, the response of $E_{tot}^N[\rho^N\{\varphi_{i,scf}^N\}]$ can be expressed as $E_{tot}^N[\rho^N] = E_{0,scf}^N + \int \Delta\rho^N(r)\theta(r,r')\Delta\rho^N(r')d^3r'd^3r$, where $\theta(r,r')$ is a density response kernel of the system. Note here we have taken advantage of that in DFT, the total energy is a function of the charge density alone, while for a given $\rho^N$, $\{\varphi_i^N\}$ can be solved through minimizing the total energy under the constraint of $\rho^N = \sum_i |\varphi_i^N|^2$. Putting all these together,

$$E_q^N = E_{0,scf}^N + \int \Delta\rho^N(r)\theta(r,r')\Delta\rho^N(r')d^3r'd^3r$$
$$+ \int \frac{\delta}{\delta\rho(r')}V_{tot}[\rho](r)\bigg|_{\rho=\rho_{0,sc}^N} \Delta\rho^N(r')\rho_d(r)d^3r'd^3r + \int \rho_d V_{tot}(\rho_{0,sc}^N)d^3r + E_{fix}[\rho_d]. \tag{19}$$

The minimum solution of $E_q^N$ in Eq. (19) in respond to $\Delta\rho^N$ is a linear equation derived from $\frac{\delta E_q^N}{\delta\Delta\rho^N(r')} = 0$:

$$2\int \theta(r',r)\Delta\rho^N(r)d^3r + \int \frac{\delta}{\delta\rho(r')}V_{tot}[\rho]\bigg|_{\rho=\rho_{0,sc}^N} \rho_d(r)d^3r = 0. \tag{20}$$

Then the charging energy $\Delta E_q^N = E_q^N - E_{0,scf}^N$ is

$$\Delta E_q^N = \frac{1}{2}\int \frac{\delta}{\delta \rho(r')}V_{tot}[\rho]\bigg|_{\rho=\rho_{0,sc}^N} \Delta\rho^N(r')\rho_d(r)d^3r'd^3r + \int \rho_d V_{tot}(\rho_{0,sc}^N)d^3r + E_{fix}[\rho_d]$$

$$= \frac{1}{2}\int \Delta\rho^N(r')\frac{1}{|r-r'|}\rho_d(r)d^3r'd^3r + \frac{1}{2}\int \frac{\delta}{\delta\rho}V_{xc}[\rho]\bigg|_{\rho=\rho_{0,sc}^N}\Delta\rho^N(r)\rho_d(r)d^3r \quad (21)$$

$$+\int \rho_d V_{tot}(\rho_{0,sc}^N)d^3r + E_{fix}[\rho_d].$$

Here we have used $V_{tot}[\rho,r] = \int \frac{\rho(r')}{|r-r'|}d^3r' + V_{xc}(\rho(r)) + V_{ion}(r)$. The above formula is a rigorous result under the second-order expansion of the total energy regard to $\rho_d(r)$ which is the charge density of the defect state wave function $\rho_d = \varphi_d^2$ under the SCF solution. To proceed, we will now use our defect charge model described in Section II.C.

According to Eqs. (9), (11), (15), there is

$$\begin{aligned}\Delta\rho^N &= \rho_{d,sc}^N - \rho_d \\ &= \rho_{d,sc}^{\infty,core} - \rho_d + C^N \\ &= \Delta\rho_{d,sc}^{\infty,core} + C^N.\end{aligned} \quad (22)$$

Plugging Eq. (22) into Eq. (21), we have

$$\Delta E_q^N = (T+E_{NL})[\varphi_d] + \frac{1}{2}\int \frac{\delta}{\delta\rho}V_{xc}[\rho]\bigg|_{\rho=\rho_{0,sc}^\infty}(\Delta\rho_{d,sc}^{\infty,core}(r)+\rho_d(r))\rho_d(r)d^3r$$

$$+\int \rho_d V_{tot}(\rho_{0,sc}^\infty)d^3r + \frac{1}{2}\int_{P,N}\rho_{d,sc}^N(r)\frac{1}{|r-r'|}\rho_d(r')d^3r d^3r' \quad (23)$$

$$+q\cdot\Delta V_N + \frac{1}{2}\int \frac{\delta}{\delta\rho}V_{xc}[\rho]\bigg|_{\rho=\rho_{0,sc}^\infty} C^N(r)\rho_d(r)d^3r.$$

Note, to derive Eq. (23), we have kept the energy to the second order of $\rho_d$ and approximated: $\frac{\delta}{\delta\rho}V_{xc}[\rho]\big|_{\rho=\rho^N} = \frac{\delta}{\delta\rho}V_{xc}[\rho]\big|_{\rho=\rho_{0,sc}^N} = \frac{\delta}{\delta\rho}V_{xc}[\rho]\big|_{\rho=\rho_{0,sc}^\infty}$. Moreover, $V_{tot}(\rho_{0,sc}^N) = V_{tot}(\rho_{0,sc}^\infty) + \Delta V_N$ and the $\Delta V_N$ is a constant. This is a good approximation since the long-range Coulomb effect vanishes for the charge-neutral system. Then the finite system potential approaches the infinite system potential quickly up to a constant (due to the $V_{tot}(G=0)=0$ requirement for the electrostatic part). The $\Delta V_N$ can be obtained by comparing the $V_{tot}(\rho_{0,sc}^N,r)$ (at $r$ far away from the defect) with bulk potential $V_{host}(r)$ of the host crystal (without defect) at the corresponding point $r$. Because, for the infinite system, the $V_{tot}(\rho_{0,sc}^\infty,r)$ at a position far away from the defect should equal to $V_{host}(r)$. It is a common trick used for defect calculations. Also note that, in Eqs.(1)-(21), all the Coulomb double integrations are done with a constant charge background to avoid the divergence as in the standard treatment of a plane wave code. In other words, in the reciprocal lattice, the reciprocal vector G=0 component of the electrostatic potential has been set to zero. Starting from Eq. (23), we have used the symbol $\int_{P,N}$ to emphasize this point, as a Coulomb double integral carried out in supercell size N with Periodic boundary condition, hence with a uniform background charge compensation. This is to be distinguished from the symbol $\int_{\infty}$ to be used later, which is the Coulomb integration when N is infinity. This $\int_{\infty}$ is also equivalent of carrying out the Coulomb integration in finite supercell N with an open boundary condition without uniform background charge compensation.

Now, to present our charge correction procedure, what we like to do is to deduce the charging energy for an infinite (thus converged) supercell $\Delta E_q^{\infty}$ from the finite supercell charging energy $\Delta E_q^N$. Note, in Eq. (23), the first line is independent of supercell size N (we can safely assume $\varphi_d$ and $\rho_d = \varphi_d^2$ are N-independent as soon as the supercell is larger than the defect wave function). As a result, in $E_C^N = \Delta E_q^{\infty} - \Delta E_q^N$, the first line in Eq. (23) will be canceled out, then we have

$$E_C^N = -q\Delta V_N + \frac{1}{2}\int_{\infty} \rho_{d,sc}^{\infty,core}(r) \frac{1}{|r-r'|}\rho_d(r')d^3rd^3r'$$
$$-\frac{1}{2}\int_{P,N} \rho_{d,sc}^{N,core}(r) \frac{1}{|r-r'|}\rho_d(r')d^3rd^3r' - \frac{1}{2}\int \frac{\delta}{\delta\rho}V_{xc}[r]\bigg|_{\rho=\rho_{0,sc}^{\infty}} C^N(r)\rho_d(r)d^3r. \quad (24)$$

In deriving Eq. (24), we have assumed $\int_{P,N} C^N(r) \frac{1}{|r-r'|}\rho(r')d^3rd^3r' = 0$ by taking the advantage that $C^N(r)$ is more or less a constant, which has been verified above in Fig. 3. The symbol $\int_{\infty}$ indicates that the integration is conducted in an infinite supercell, not in the periodic finite supercell with size N. In the actual calculation, this integration can be realized by solving the Poisson equation in an open boundary condition. We have used a technique of applying a double size supercell and truncate the Coulomb interaction range beyond the original supercell size. Moreover, we apply FFT to calculate the Poisson equation in the open boundary condition. This technique has been implemented in the plane-wave material simulations (PWmat) code [32] (see supplemental information section 9 for detailed explanation). Tests show that the last term in Eq. (24) is rather small, so we also can ignore it. We finally have our image charge correction formula as:

$$E_{C-SC}^{N} = -q\Delta V_{N} + \frac{1}{2}\int_{\infty} \rho_{d,sc}^{\infty,core}(r)\frac{1}{|r-r'|}\rho_{d}(r')d^{3}rd^{3}r'$$
$$-\frac{1}{2}\int_{P,N} \rho_{d,sc}^{\infty,core}(r)\frac{1}{|r-r'|}\rho_{d}(r')d^{3}rd^{3}r'. \quad (25)$$

In our scheme, the procedure of the image charge correction contains three steps after the conventional defect calculation at supercell size N: First, get the potential alignment (PA) between the local potential of the supercell with the neutral defect and the pristine crystal bulk potential, i.e., $\Delta V_N = V_{tot}(\rho_{0,sc}^N, r) - V_{host}(r)$ at $r$ far away from the defect; Second, use Eq. (13) to calculate $\rho_{d,sc}^{N,core}(r)$ as an approximation of $\rho_{d,sc}^{\infty,core}(r)$ to be used in Eq. (25); Third, utilize the q-charged defect SCF result to get the defect wave function $\varphi_d$ and the charge density $\rho_d = \varphi_d^2$ to be used in Eq. (25). After these three steps, Eq. (25) is ready to gain $E_{C-SC}^N$. The most significant point of our scheme is the use of $\rho_{d,sc}^{\infty,core}(r)$ which corresponds to a screened charge density of $\rho_d$. In a sense, Eq. (8) is like to approximate $\rho_{d,sc}^{\infty,core}(r)$ as $\rho_d(r)/\varepsilon$. However, applying bulk dielectric constant ε might be inappropriate due to the small defect size, local field effect, or even nonlinear response near the defect. On the other hand, as we show here, the information for dielectric response is already included in the SCF calculation and the resulting $\rho_{d,sc}^{\infty,core}(r)$. One drawback of our method is that, in order to use Eqs. (11)-(13) to obtain $\rho_{d,sc}^{\infty,core}(r)$, we have to carry out calculations with at least two supercells with different sizes.

In our derivation above, we have expanded the energy up to the second order of $\rho_d$. Strictly speaking, this is only valid for linear response theory, thus cannot be used for cases where nonlinear screening effect exists. However, our defect screening model is general and is not restricted to the linear screening case. There is a subtle difference between the screening close to the defect versus the screening far away from the defect. We expect any nonlinear screening effect will only happen near the defect (where the electric field is strong, e.g., represented by the total charge of $\rho_{d,sc}^{N,core}(r)$), which can be captured by the actual SCF calculation and the defect screening model. On the other hand, our final image correction energy (which is an energy deviation between the finite supercell and the infinite-size supercell) only concerns the screening far away from the defect (which is always linear). And the nonlinear screening energy near the defect should be canceled out between the finite supercell and infinite supercell energies. This might validate Eq. (25) to be used even for nonlinear screening cases (since the inaccurate part of the second-order expansion in its derivation near the defect should nevertheless be canceled out between the finite supercell expression and infinite supercell expression). On the other hand, the nonlinear screening effects (e.g., the total charge of $\rho_{d,sc}^{N,core}(r)$) can be described by the SCF calculation and the defect screening model, and then be used in Eq. (25).

It is also worth discussing the ionic screening effect here. It is likely that the Eq. (25) is also valid, even including ionic screening. First, Eq. (16) is still correct, although in Eq. (17) we should include the ion-ion Ewald interaction term. The perturbation induced by the interaction between $\rho_d$ and the ionic charge can still be written down as dot product between $\rho_d$ and the ionic displacement (or say the ionic screening charge as in Eq. (18)). The total energy cost by ionic displacement will also have a second-order harmonic oscillator form of the ionic displacement, much like in $E_{tot}^N[\rho] = E_{0,scf}^N + \int \Delta\rho^N(r)\theta(r,r')\Delta\rho^N(r')d^3r'd^3r$. For the minimum solution, this cost will still equal to $-1/2$ of the linear interaction term in Eq. (20). Thus, the whole derivation can still go through. We will have the same Eq. (25) if the screening model also holds for ionic screening, but, the $\rho_{d,sc}^{\infty,core}$ should include the charge response from the ionic displacement. Nevertheless, it needs to be tested in the future whether the ionic screening also follows our defect charge screening model. We will defer the investigation for the ionic screening effect in the future study. In the following discussions, we will restraint ourselves in cases where the ionic screening effect is small, or we deliberately ignore the ionic screening contribution. Correspondingly, for the models where the bulk dielectric constant ε has to be used in FNV and C-AP schemes, we always use $\varepsilon_\infty$. To simplify the comparison, we also use the neutral charge atomic positions for the charged-defect calculation.

We note that, the image correction in Eq. (25) is similar to the FNV method where the interaction between the bare defect charge and the screened potential is used to describe the image correction term. The difference is that, in the FNV, an analytical model, more specifically a dielectric constant is used to yield the screened potential from the bare charge, and in our current method, the screened potential is obtained from the screened charge, which is obtained directly from DFT supercell defect calculation, thus no dielectric constant approximation is necessary.

### E. Comparison with the FNV method

In FNV scheme, the defect-induced potential is divided into short-range and long-range potentials. The long-range potential is produced by a defect charge model $q^{model}$ and results in $-E^{lat}[q^{model}]$, and the short-range potential induces energy $q\Delta v$. A general analytical model is used to describe $\rho_d(r)$ in Eq. (8) and has a spherical symmetric form,

$$\rho_d(r) = qxN_\gamma e^{-r/\gamma} + q(1-x)N_\beta e^{-x^2/\beta^2}. \tag{26}$$

$N_\gamma$ and $N_\beta$ denote the normalization constants for exponential and the Gaussian terms, $\beta$ determines the width of Gaussian charge, and the decay constant $\gamma$ and the tail weight $x$ are obtained by fitting the defect wave functions [17]. After getting the $q^{model}$ and the long-rang potential, the defect-induced potential (from DFT calculations of neutral and q-charged defect states) subtracting the long-rang part is the short-range part. If the $q^{model}$ is appropriate, the short-range potential will reach a plateau value of $-\Delta v$ at $r$ far away from the defect. The plateau gives rise to the short-range correction $q\Delta v$. Then the total correction to the energy is $-E^{lat}[q^{model}] + q\Delta v$. Here, q is the defect charge state, not the number of ionized electrons as in our scheme.

This $q^{model}$ fitting can be complicated, especially when the defect wave function is anisotropic. One can imagine that for more complex defects, e.g., a two-atom center co-doping defect, the $\rho_d(r)$ can be nonspherical, which causes some uncertainty in using Eq. (26). One uncertainty of the FNV scheme is the plateau value $-\Delta v$. The plateau might depend on the orientation of the planar-averaged potential. For example, the planar-averaged electrostatic potential along different lattice vector directions in relaxed β-Ga$_2$O$_3$ supercells is apparently different, as shown in Fig. S15 in the supplemental information. β-Ga$_2$O$_3$ has $C2/m$ symmetry, and the potential along the lattice vector $a$ (the lattice vector with the longest lattice constant) fluctuates significantly, showing no plateau, presented in Fig. S15(a). We found this increases the uncertainty when calculating the correction energy in our implementation with FNV.

Actually FNV and C-AP are equavelent except the origin of the defect charge model. FNV need a q model while C-AP formulated in Eq. (8) with a further model approximation for the defect charge density $\rho_d(r)$. In the following tests, FNV correction energy calculated using the code Sxdefectalign [33] provided by Christoph Freysoldt is compared with C-AP method formulated in Eq. (8) using the same $\varepsilon_\infty$ but with explicit $\rho_d$ calculated from Eq. (3). For most cases, the correction energies provided by FNV and C-AP are very close. Only for a few defects like $V_C^{2+}$ in diamond, $V_O^{2+}$ in MgO, a small deviation (about 0.2 eV) arises due to the use of the analytical charge model in FNV method.

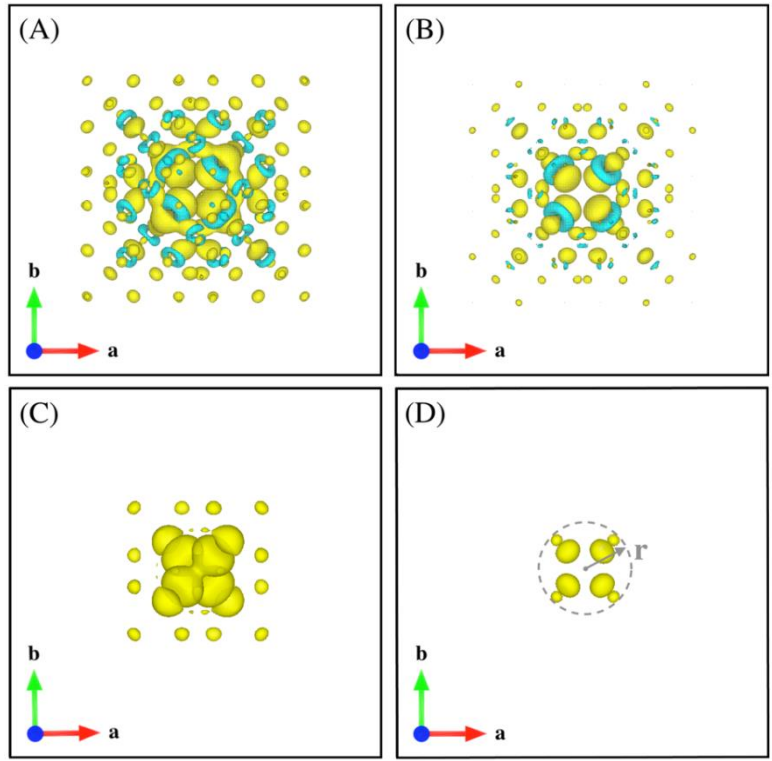

FIG.4. Defect charge distribution for $V_C^+$ in a cubic 512-atom diamond supercell (supercell length is a = 14.13 Å). (A) shows the screened defect charge density $\rho_{d,sc}^N$, (B) core defect charge density $\rho_{d,sc}^{N,core}$, (C) unscreened defect charge density $\rho_d$, and (D) $\rho_d/\varepsilon_\infty$ ($\varepsilon_\infty = 5.62$). All plots view normal to the (001) plane with an isosurface value of $\pm 0.002$ (e/Bohr$^3$). The $V_C^+$ defect is located at the center of the supercell. The black box indicates the boundary of the 512-atom supercell.

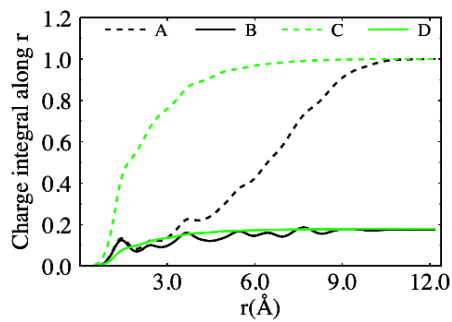

FIG.5. Charge integrations along radial direction for defect charge distributions of $V_C^+$ in 512-atom diamond supercell: (A) $\rho_{d,sc}^N$; (B) $\rho_{d,sc}^{N,core}$; (C) $\rho_d$; and (D) $\rho_d/\varepsilon_\infty$ ($\varepsilon_\infty = 5.62$). In this case, the $\rho_{d,sc}^{N,core}$ and $\rho_d/\varepsilon_\infty$ have been integrated into the same total charge. Nevertheless, this is just accidental, not always true, as illustrated in other defect cases.

As for C-SC in Eq. (25), the defect charge $\rho_{d,sc}^{N,core}$ is even more different. The $\rho_{d,sc}^N$, $\rho_{d,sc}^{N,core}$, $\rho_d(r)$, $\rho_d(r)/\varepsilon_\infty$ for a $V_C^+$ defect in a 512-atom diamond supercell are shown in Fig.4. Fig. 4(A) shows that the charge difference $\rho_{d,sc}^N$ is a little dispersive due to the existence of the background charge $C^N(r)$, while the core part $\rho_{d,sc}^{N,core}$ in Fig. 4(B) is more localized when the background part is removed. Compared with $\rho_{d,sc}^N$ and $\rho_{d,sc}^{N,core}$, the unscreened defect charge density $\rho_d$ is confined near the defect center. After screening with $\varepsilon_\infty = 5.62$ (as calculated by DFT using the same pseudopotential, etc.), the total charge reduces to $q = 0.176$ and the isosurface is confined in the range of $r \approx 2\text{Å}$ (shown in Fig. 4(D)) when the isosurface value of $\pm 0.002$ (e/ Bohr$^3$) is used. In this respect, the macroscopic dielectric constant overestimates the screening response. We integrated the defect charge density shown in Fig. 4 along the radial direction, and the corresponding results are shown in Fig. 5. We find that the screening effect reduces the core charge by 82.4%. Surprisingly, $\int \rho_{d,sc}^{N,core} 4\pi r^2 dr$ (line B in Fig. 5) and $\int \rho_d/\varepsilon_\infty 4\pi r^2 dr$ (line D in Fig. 5) seem to reach the same limit, e.g., their total amount of screening charges are the same. However, this is just a coincidence, and it is not true for other defect systems. Furthermore, Figs. 4 (B) and (D) exhibit that the charge distributions of the core defect charge density $\rho_{d,sc}^{N,core}$ and $\rho_d/\varepsilon_\infty$ in the supercell are entirely different. This substantial difference will influence their Coulomb interaction with the defect charge $\rho_d$.

In both simplified schemes formulated in Eq. (8) and FNV method, the most significant approximation might come from the use of the bulk macroscopic dielectric constant to describe the screening effect. As a result, as will be shown in our later Section, there can be substantial differences in correction energies between the FNV scheme and our final scheme in Eq. (25). This is particularly true for $+2$ charge-state defect where the image charge correction term becomes very large.

### F. Comparison with MP and LZ corrections

Makov and Payne (MP) [15] proposed an image interaction correction based on approximating the defect charge density as a sum of a point-like charge density and a more extended part $\rho_e(r)$ (where the net charge of $\rho_e(r)$ is zero). The point charge to point charge interaction results in a $1/L$ term, while the point charge to $\rho_e(r)$ interaction results in the $1/L^3$ term. Thus, we have

$$\Delta E_C = \frac{q^2 \alpha_M}{2\varepsilon L} + \frac{2\pi q Q_r}{3\varepsilon L^3}. \tag{27}$$

Here, $\alpha_M$ is a structure factor of the Madelung energy for a respective supercell geometry, $\varepsilon$ is the macroscopic dielectric constant, and $Q_r$ is the second radial moment of $\rho_e(r)$:

$$Q_r = \int_\Omega d^3 r \rho_e r^2. \tag{28}$$

This formula has been carefully tested by Lany and Zunger [8] and found to work reasonably well. Nevertheless, there are some conceptual issues. For example, in contrast to what the MP stated in their original paper, where $\rho_e(r)$ should be obtained from the unscreened charge density (e.g., in our case, the $\rho_d(r)$ minus the delta point charge, which does not contribute to the second radial moment of Eq. (28)), it is found that the SCF charge (which already includes the screening effect) needs to be used for $\rho_e(r)$. But that presents a conceptual problem for Eq. (27), since the screening effect should already be included by the dielectric constant $\varepsilon$. If $\rho_e(r)$, i.e. $Q_r$ also includes the screening effect, then there will be double counting of that effect in Eq. (27).

Some more careful considerations are provided by Lany and Zunger [7,8] to analyze the screening effect and the screening charge, leading them to present the following formula for the image correction:

$$E_C = \left[1 + c_{sh}(1-\varepsilon^{-1})\right] \frac{q^2 \alpha_M}{2\varepsilon L}. \tag{29}$$

Here $c_{sh}$ is a shape factor depending on $Q_r$. Note that, this formula contains a $1/\varepsilon^2$ term, which usually does not exist in a screening model.

Much like in FNV, one advantage of these methods is that one supercell calculation should be enough to get the image correction. In contrast, in our method, we need at least two calculations with different supercell sizes in order to yield $\rho_{d,sc}^{N,core}$. Nevertheless, the use of bulk dielectric constant $\varepsilon$ makes the above methods potentially less accurate.

## G. Effective dielectric constant for the defect screening

To test the effect of defect dielectric screening, and compare that to the macroscopic bulk dielectric constant, we can define an effective dielectric constant (or defect dielectric constant, $\varepsilon_d^N$) for defect screening. There could be many ways to define that. For us, we can define it by calculating the total screening charge. In a macroscopic picture, the screening charge for a charge $q$ will be $q_{sc}^N = \left(1 - \frac{1}{\varepsilon}\right)q$. In our case, this screening charge is just the sum of $C^N(r)$. Thus, we have

$$q_{sc}^N = \int C^N(r) d^3 r, \tag{30}$$

and then

$$\varepsilon_d^N = \frac{q}{q - q_{sc}^N}. \tag{31}$$

The calculated effective dielectric constants for $V_O^+$ and $V_O^{2+}$ in MgO at different supercell sizes are listed in Table I. In comparison with the PBE calculated host bulk dielectric constants of $\varepsilon_\infty = 3.14$ (we have used a long slab method to calculate these bulk dielectric constants for all the host materials, which agree well with literature results) for MgO. The effective dielectric constants in the table are all remarkably larger than bulk $\varepsilon_\infty$. For supercells with $V_O^{2+}$, $\varepsilon_d^N$ has the largest value as 7.18 for MgO, respectively, which means the screening is much stronger and nonlinear (as the dielectric constant increases with the defect charge).

Table I. Effective dielectric constant $\varepsilon_d^N$ of $V_O^+$ and $V_O^{2+}$ defects in MgO with different supercell size N calculated according to Eq. (31). For comparison, the PBE calculated bulk macroscopic dielectric constants $\varepsilon_\infty$ of host MgO is presented together.

| Defects | supercell | | | | $\varepsilon_\infty$ |
|---|---|---|---|---|---|
| | 64 | 216 | 512 | 1000 | |
| $V_O^+$ | 3.31 | 3.38 | 3.39 | 3.42 | 3.14 |
| $V_O^{2+}$ | 4.10 | 4.88 | 6.23 | 7.18 | |

## III. APPLICATIONS TO DEFECTS

### A. Calculation details

We have calculated the formation energy of different defects in various host materials having relatively wide band gaps, including oxides for which image charge correction can be very considerable. These defects include: $V_O^+$ and $V_O^{2+}$ in MgO; $V_C^+$, $V_C^{2+}$, $NV^-$ [34,35], and $SiV^-$ [36] in diamond; $V_{Ga}^{3-}$ in GaAs; a complex defect ($V_{Ga} - O_N^-$) in GaN [37]; as well as $V_{Ga}^-$ and $V_{Ga}^{2-}$ in β-Ga$_2$O$_3$; $V_O^+$ and $V_O^{2+}$ in ZnO.

Here we summarize the computational details. All the calculations are performed with PWmat [32] package using the SG15 collection of the Optimized Norm-Conserving Vanderbilt Pseudopotentials (ONCV) [38]. We adopt local density approximation (LDA) for diamond and GaAs as in Ref [20,28] and Perdew-Burke-Ernzerhof of Generalized Gradient Approximation (GGA-PBE) for MgO, ZnO, GaN, and β-Ga$_2$O$_3$. To assist with the convergence comparison, we have used an equivalent Monkhorst-Pack k-point mesh for different supercell sizes. They are all equivalent to have 3×3×3 k-points for a 512-atom supercell (thus smaller supercell will have more k-pints). The primitive cell is well optimized, as shown in Table II.

Table II. Calculated and experimental lattice constants and ion-clamped dielectric constants (electronic dielectric constant, $\varepsilon_\infty$) of pristine bulk. (The **bold** numbers are used as $\varepsilon_\infty$ in FNV/C-AP). The space group for MgO, diamond, GaAs, GaN, β-Ga$_2$O$_3$ and ZnO are $Fm\bar{3}m$, $Fd3m$, $F\bar{4}3m$, $P6_3mc$, $C2/m$ and $P6_3mc$, respectively. The values in the rows of Theory and Expt. are taken from literature for comparison. The first row values for each material are calculated in this work.

| Host | Lattice constant(Å) | $\varepsilon_\infty$ | Functional |
|---|---|---|---|
| MgO | 4.22 | **3.14** | GGA-PBE |
| Theory [a] | 4.25 | 3.16 | |
| Expt. [b] | 4.207 | 3.0 | |
| Diamond | 3.53 | **5.62** | LDA |
| Theory [c] | 3.536 | 5.76 | |
| Expt. [d] | 3.567 | 5.7 | |
| GaAs | 5.59 | **12.78** | LDA |
| Theory [e] | 5.627 | 13.7 | |
| Expt. [f] | 5.642 | 11.1 | |
| GaN | 3.23/5.26 | **6.10**($\varepsilon_\parallel$) | GGA-PBE |
| Theory [g] | 3.22/5.22 | 5.60($\varepsilon_\parallel$) 5.54($\varepsilon_\perp$) | |
| Expt. [h] | 3.198/5.182 | 5.37($\bar{\varepsilon}_\infty$) | |
| β-Ga$_2$O$_3$ | 12.37/3.06/5.68 | **3.92**($\varepsilon_\parallel$) | GGA-PBE |
| Theory [i] | 12.446/3.083/5.876 | 3.55($\bar{\varepsilon}_\infty$) | |
| Expt. [j] | 12.214/3.037/5.798 | 3.57($\bar{\varepsilon}_\infty$) | |
| ZnO | 3.24/5.21 | **5.38**($\varepsilon_\parallel$) | GGA-PBE |
| Theory [k] | 3.286/5.299 | 5.20($\varepsilon_\parallel$) 5.22($\varepsilon_\perp$) | |
| Expt. [l] | 3.250/5.207 | 3.70($\varepsilon_\parallel$) 3.78($\varepsilon_\perp$) | |

[a]Reference [20]

[bce]Reference [28]

[d]Reference [39]

[f]Reference [40]

[g]Reference [41,42]

[h]Reference [43,44]

[i]Reference [45,46]

[j]Reference [47,48]

[k]Reference [20,49]

[l]Reference [50,51]

## B. The convergence of C-SC, C-AP, FNV corrected formation energies

Four different methods are used to provide the image charge corrections: NC denoting no image charge correction, FNV formulated in Eq. (26), C-AP (correction using an approximation of macroscopic dielectric constant) in Eq. (8) and C-SC (correction scheme with screened charge density model) in Eq. (25). According to the correction effects, we separate these 12 types of defects into two groups. One group includes $V_O^+$ and $V_O^{2+}$ in MgO. The other includes the rest of the defects.

**Energy convergence for defects in MgO**

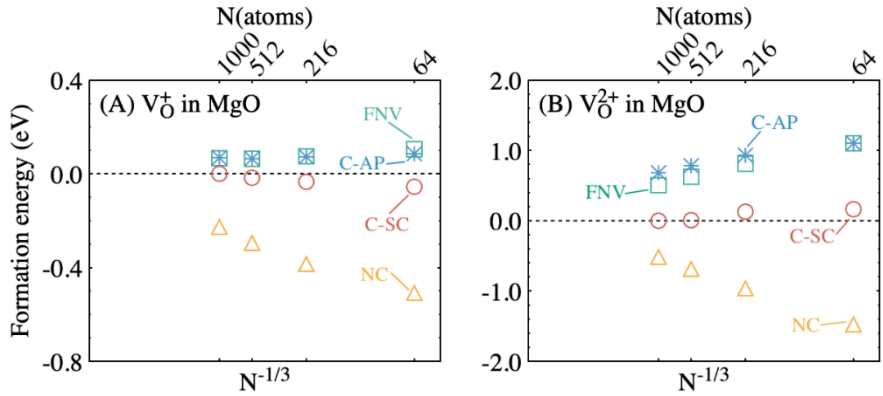

FIG.6. Corrected formation energies of (A) $V_O^+$ and (B) $V_O^{2+}$ defects in MgO with corrections based on NC, FNV, C-AP, and C-SC schemes. $\varepsilon_\infty = 3.14$ are used for MgO, in FNV and C-AP corrections. All corrected formation energies are referenced to the C-SC corrected formation energy (which is set to zero) at the largest calculated supercell. The numbers at the top of the panels indicate the numbers of atoms contained in the supercell. The left edge of the box correspond to the position when N is infinity.

Fig. 6 shows formation energies of $V_O^+$ and $V_O^{2+}$ defects in MgO corrected by NC, FNV, C-AP, and C-SC schemes. For $V_O^+$ and $V_O^{2+}$ in MgO supercells, the image charge correction is quite significant with more than 0.2~0.5 eV for $V_O^+$ and 0.5~1.5 eV for $V_O^{2+}$ (the range corresponds to different supercell sizes: the ampler supercell causes a smaller correction as expected). The formation energies corrected by C-SC scheme are converged the fastest concerning supercell size for the defects in this group. However, FNV/C-AP and C-SC schemes give distinct results. FNV and C-AP always overestimate the image correction. As expected, the correction energies for +2 defect charge states are about 4 times the correction energies for +1 defect charge state. As discussed in II.E, it is inaccurate to describe the screening effect with a bulk macroscopic dielectric constant from the host material. The effective dielectric constants $\varepsilon_d^N$ of various MgO supercells with $V_O^+$ and $V_O^{2+}$ defects calculated by Eq. (31) are listed in Table I. They are all larger than the electronic dielectric constants $\varepsilon_\infty$ of bulk, 3.14 for MgO. With the incresing supercells, $\varepsilon_d^N$ of MgO for $V_O^+$, $V_O^{2+}$ defect is incresing. For 1000-atom supercell, suprisingly, $\varepsilon_d^N$ is almost twice of the bulk value 3.14. That explains why FNV and C-AP schemes, which utilize bulk macroscopic dielectric constant for screening effect, provide an overestimation of the correction energy. Note that, these values contain no ionic contribution since the atomic positions are deliberately fixed at their atom positions of neutral charge state. It is unlikely the enhanced $\varepsilon_d^N$ value is attributed to the finite supercell size (large reciprocal space vector q) effect or local field effect since they usually lead to smaller effective screening. The fact that +2 defect has a much larger dielectric constant than +1 defect leads us to believe that nonlinear screening effect plays an important role. Considering the large additional charge density at the defect and the strong electric field for a charged defect, it is not surprising that the dielectric screening is nonlinear. Such a nonlinear screening effect can be challenging to model analytically, pointing to a potential challenge in using the analytical models.

**Energy convergence of other defects**

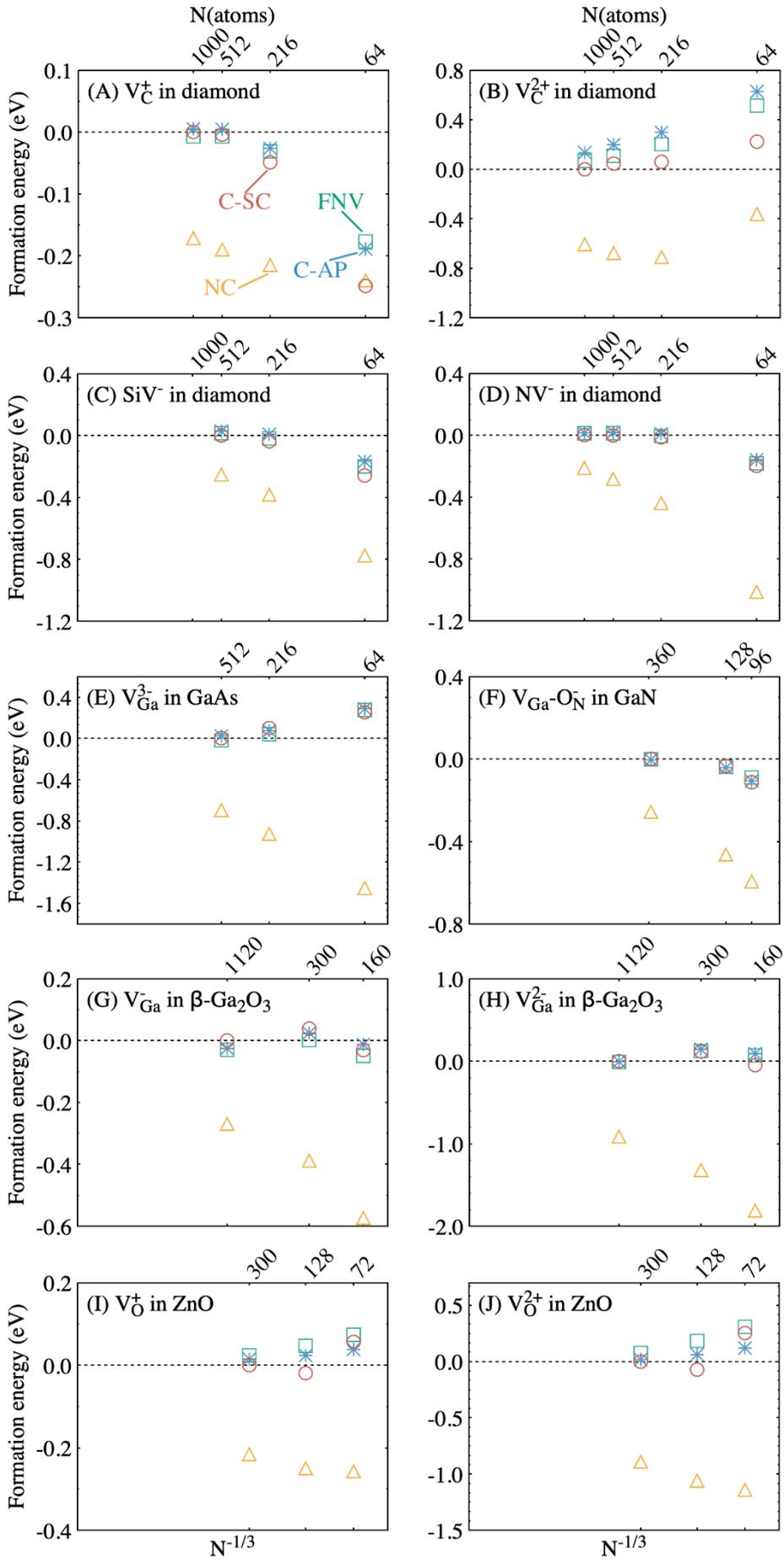

FIG.7. Corrected formation energies of (A) $V_C^+$, (B) $V_C^{2+}$, (C) $SiV^-$, and (D) $NV^-$ defects in diamond, (E) $V_{Ga}^{3-}$ in GaAs, (F) $V_{Ga}-O_N^-$ in GaN, (G) $V_{Ga}^-$ and (H) $V_{Ga}^{2-}$ in β-Ga2O3, (I) $V_O^+$ and (J) $V_O^{2+}$ in ZnO with corrections based on NC, FNV, C-AP, and C-SC schemes. Related $\varepsilon_\infty$ used in FNV and C-AP schemes are listed in Table II (the bold numbers) and Table III. All corrected formation energies are referenced to the C-SC corrected formation energy (which is set to zero) at the largest calculated supercell for each material. The numbers at the top of the panels indicate the number of atoms contained in the supercell. The left edge of the box correspond to the position when N is infinity.

Fig. 7 shows formation energy convergence of diamond supercells containing $V_C^+$, $V_C^{2+}$, $NV^-$, and $SiV^-$ defects, respectively; GaAs supercells having a $V_{Ga}^{3-}$; GaN supercells with a $V_{Ga}-O_N^-$ complex defect; β-Ga₂O₃ with a $V_{Ga}^-$ or $V_{Ga}^{2-}$ defect; ZnO with a $V_O^+$ or $V_O^{2+}$. From Fig. 7, we can see that FNV, C-SC, and C-AP schemes work equally well, and they all converge quickly even for the highly charged defect $V_{Ga}^{-3}$. One of the reasons for their similarity is the fact that their effective defect dielectric constants $\varepsilon_d^N$ are quite close to the bulk dielectric constant $\varepsilon_\infty$. Table III lists the calculated effective defect dielectric constants from Eq. (31) using the largest supercell, and comparing these results with the bulk macroscopic dielectric constants calculated using the same pseudopotential and exchange-correlation functional. As shown in Table III, the $\varepsilon_d^N$ for $V_C^+, V_C^{2+}, NV^-$, and $SiV^-$ defects in diamond is 5.60, 6.54, 6.25, and 6.64, respectively. They are all close to the bulk diamond dielectric constant of $\varepsilon_\infty = 5.62$. It is interesting to note that, there is no sizeable nonlinear effect in this case, as the $\varepsilon_d^N$ from $V_C^+$ to $V_C^{2+}$ only changed from 5.60 to 6.54. The situations for GaAs, GaN, β-Ga₂O₃ and ZnO are also similar. Their $\varepsilon_\infty$ are 12.78, 6.10, 3.92, 5.65 respectively, all close to their defect effective $\varepsilon_d^N$. At this stage, we have found no prior way to guess which system will have a sizeable nonlinear screening effect, and which system will have no such effect. As a result, the direct numerical calculation is the only reliable way to find this out.

Table III. Defect dielectric constants $\varepsilon_d^N$ (Eq. (31)) of different defects compare with the macroscopic dielectric constants $\varepsilon_\infty$ of pristine bulk. For different defects, $\varepsilon_d^N$ is from the supercell with the largest size. Note, $\varepsilon_\infty$ is not $\varepsilon_d^{N=\infty}$, instead it is a bulk macroscopic dielectric constant. Due to local field effect and finite length effect (finite reciprocal vector q in the screening dielectric constant), as well as possible nonlinear effect, near the defect, the $\varepsilon_d^{N=\infty}$ can be different from $\varepsilon_\infty$.

| Host | Defect | $\varepsilon_d^N$ | $\varepsilon_\infty$ |
|---|---|---|---|
| MgO | $V_O^+$ | 3.42 | 3.14 |
|  | $V_O^{2+}$ | 7.18 |  |
| Diamond | $V_C^+$ | 5.60 | 5.62 |
|  | $V_C^{2+}$ | 6.54 |  |
|  | $NV^-$ | 6.25 |  |
|  | $SiV^-$ | 6.64 |  |
| GaAs | $V_{Ga}^{3-}$ | 12.91 | 12.78 |
| GaN | $V_{Ga} - O_N^-$ | 5.69 | 6.10 |
| β-Ga$_2$O$_3$ | $V_{Ga}^-$ | 3.37 | 3.92 |
|  | $V_{Ga}^{2-}$ | 3.74 |  |
| ZnO | $V_O^+$ | 4.90 | 5.38 |
|  | $V_O^{2+}$ | 5.65 |  |

## IV. CONCLUSIONS

In summary, we provide a rigorous derivation for the image interaction correction formula based on a defect charge screening model. This charge screening model is tested via numerical calculations. In this model, the screened charge of the defect is separated into a screened core charge $\rho_{d,sc}^{N,core}(r)$ and a close to constant (especially when away from the defect) background compensation charge $C^N(r)$. While the core charge approaches $\rho_{d,sc}^{\infty,core}(r)$ quickly, the background charge $C^N(r)$ is inversely proportional to the supercell volume. An image interaction correction is provided by the difference of the Coulomb interaction energies between $\rho_{d,sc}^{\infty,core}(r)$ and the bare defect charge $\rho_d(r)$, calculated in periodic supercell and infinite system, respectively (Eq. (25)). We believe Eq. (25) is also valid when ionic screening is also included, though we have excluded the ionic screening contribution in our tests currently. We also argue that Eq. (25) can be applied to the cases where the nonlinear screening effect is important. Such a nonlinear screening effect can be captured by the SCF calculations and represented by the defect screening model as exemplified by the total charge of $\rho_{d,sc}^{\infty,core}(r)$. Using the screening charge from $C^N(r)$, it is possible to define an effective dielectric constant for the defect (Eq. (31)). We found that, in the cases of MgO, the defect dielectric constant is much larger than the macroscopic dielectric constant. We attribute this to the nonlinear screening effect. We also found that, in such case, our image interaction correction results are very different from the results of the previous method like the FNV method. There are, however, also other cases where the defect dielectric constant is close to the macroscopic dielectric constant. In those cases, especially for single charged defect and high symmetry defects, our image interaction correction results are similar to previous method results. Our approach is different from previous methods in that it uses additional numerical calculations to figure out the image interaction correction term, instead of using simplified analytical models. There is no need to use the macroscopic dielectric constant, as the SCF calculation has already captured the screening effect. Especially, our method can be used to calculate the vertical transition energy which has been studied with different methods [52,53] very recently. For vertical transition, the atoms are fixed during the ionization of defect so there is no ionic screening, which can be studied within our derivation and leave us to do a further work in the future.

## ACKNOWLEDGMENTS

L.W. W was supported by the Director, Office of Science (SC), Basic Energy Science (BES), Materials Science and Engineering Division (MSED), of the US Department of Energy (DOE) under Contract No. DE-AC02-05CH11231 through the Materials Theory program (KC2301). The work in China was supported by the National Natural Science Foundation of China (NSFC) under Grant Nos. 11925407 and 61927901.